\documentclass[letterpaper, 10pt, conference]{ieeeconf}  

\IEEEoverridecommandlockouts                              

\usepackage[table]{xcolor}
\usepackage{cite}
\usepackage{amsmath,amssymb,amsfonts}
\usepackage{algorithmic}
\usepackage{graphicx}
\usepackage{algorithm,algorithmic}
\usepackage{hyperref}
\usepackage{booktabs} 
\usepackage{subcaption}
\hypersetup{hidelinks}
\usepackage{textcomp}
\usepackage{xcolor}
\usepackage{tcolorbox}
\usepackage{xspace}
\usepackage{comment}
\usepackage{tikz}
\usepackage{pifont}
\usepackage{pgfplots}
\pgfplotsset{compat=1.18}
\usepgfplotslibrary{colorbrewer}
\usepackage{csvsimple}
\usepackage{makecell}

\newtheorem{definition}{Definition}

\newcommand{\mname}[0]{\textsc{EpiCAge}\xspace}

\def\BibTeX{{\rm B\kern-.05em{\sc i\kern-.025em b}\kern-.08em
    T\kern-.1667em\lower.7ex\hbox{E}\kern-.125emX}}
\title{Integrating Epigenetic and Phenotypic Features for Biological Age Estimation in Cancer Patients via Multimodal Learning}
\author{
Shuyue Jiang$^{1}$, Wenjing Ma$^{2}$, Shaojun Yu$^{1}$, Chang Su$^{1}$, Runze Yan$^{1}$, and Jiaying Lu$^{1}$\textsuperscript{\ding{41}}
\thanks{$^{1}$Emory University, Atlanta, GA 30322, USA.}
\thanks{$^{2}$University of Michigan, Ann Arbor, MI 48109, USA.}
\thanks{\textsuperscript{\ding{41}}Corresponding Author: Jiaying Lu, Ph.D. (jiaying.lu@emory.edu)}
}

\begin{document}
\maketitle

\begin{abstract}
Biological age, which may be older or younger than chronological age due to factors such as genetic predisposition, environmental exposures, serves as a meaningful biomarker of aging processes and can inform risk stratification, treatment planning, and survivorship care in cancer patients. We propose \mname, a multimodal framework that integrates epigenetic and phenotypic data to improve biological age prediction. Evaluated on eight internal and four external cancer cohorts, \mname consistently outperforms existing epigenetic and phenotypic age clocks. Our analyses show that \mname identifies biologically relevant markers, and its derived age acceleration is significantly associated with mortality risk. These results highlight \mname as a promising multimodal machine learning tool for biological age assessment in oncology.

\end{abstract}

\section{Introduction}

Biological age estimation provides a quantitative measure of an individual’s physiological state, which may deviate from their chronological age due to genetic, environmental, and disease-related factors~\cite{lopez2013hallmarks}. Unlike chronological age, biological age aims to capture the cumulative burden of aging processes at the molecular and cellular levels, offering a more precise indicator of health status and disease susceptibility. In cancer patients, biological age estimation is particularly valuable, as both the malignancy itself and its treatments can significantly accelerate biological aging~\cite{levine2018epigenetic}. Previous studies have shown that increased biological age is associated with worse clinical outcomes, including higher mortality, reduced treatment tolerance, and impaired functional recovery~\cite{yu2020epigenetic,dugue2020epigenetic}. Therefore, accurate estimation of biological age holds promise for improving treatment personalization, and long-term survivorship care in cancer populations.

Researchers have developed various biological age estimation models, which are commonly referred to as clock models, to quantify physiological aging across individuals. These models are typically trained to predict an individual’s chronological age from biological features (e.g., transcriptomic, epigenetic or phenotypic markers)~\cite{horvath2013dna,peters2015transcriptional}, under the assumption that deviations between predicted and actual age reflect underlying biological aging processes.
Among various types of clock models, \textit{epigenetic clocks} utilize DNA methylation (DNAm) data to measure aging-related molecuar changes.
Epigenetic alterations, one of the nine aging hallmarks~\cite{lopez2013hallmarks}, can be precisely quantified through DNAm profiling, enabling the estimation of biological age by analyzing methylation levels at specific CpG sites across the genome~\cite{horvath2013dna,hannum2013genome}.
In parallel, \textit{phenotypic clocks} offer an alternative approach to biological age estimation by utilizing easily accessible clinical and physiological indicators such as blood pressure, inflammatory biomarkers, grip strength, and walking speed~\cite{wagner2016biomarkers,nofuji2016associations}. Both epigenetic and phenotypic clock models demonstrate remarkable accuracy in predicting chronological age and show strong associations with age-related outcomes such as morbidity and mortality~\cite{levine2018epigenetic,warner2024systematic}.

Given the complementary nature of DNA methylation (DNAm) and clinical variables for aging, we propose to integrate both data types to build a more comprehensive model for biological age estimation. DNAm captures molecular signatures of cellular aging, while clinical features reflect systemic physiological and functional status. By combining these modalities, our \mname model leverages the molecular precision of epigenetic clocks alongside the accessibility and interpretability of phenotypic markers.
Specifically, \mname is designed as a multi-layer stacked multimodal framework. In the first layer, it includes three parallel clock models: an epigenetic clock that is based solely on DNAm features, a phenotypic clock that relies on clinical-pathological variables, and a fusion clock that integrates both data. The second layer is a higher-level fusion clock that ensembles predictions from the first layer, further enhanced by a skip connection~\cite{he2016deep} to the original input features. The architecture is modular, supporting flexible implementation choices; we implement two variants using either the lightweight ElasticNet~\cite{zou2005elasticnet} or the large-scale foundation model TabPFN~\cite{hollmann2025tabpfn}.
This multimodal strategy enhances the robustness and clinical utility of biological age prediction, particularly in cancer populations where tumors and treatments can uniquely alter both molecular and functional aging trajectories.

In this study, we curate two datasets from the publicly available multi-omics resource LinkedOmics~\cite{vasaikar2018linkedomics}, which provides both epigenetic and phenotypic data from cancer patients. Eight cancer types are selected to form the internal dataset for training and validation, while four distinct cancer types are used to construct the external dataset for evaluating the generalizability of our model across independent cohorts.
\mname consistently outperforms existing epigenetic clocks~\cite{horvath2013dna,hannum2013genome} and phenotypic models~\cite{chen2016xgboost,erickson2020autogluon}, achieving lower prediction errors on both internal and external datasets. We further investigate the biological relevance of the CpG sites identified by \mname using evidence-based analyses such as KEGG pathway enrichment~\cite{garcia2022functional}, revealing that many selected CpGs are distinct from those used in traditional clocks and are highly associated with aging and cancer. To evaluate clinical utility, we conduct survival analysis relating \mname-derived age acceleration to 5-year mortality risk, and observe a significant positive association—supporting the prognostic value of our model. Finally, we perform error analysis and ablation studies to assess the robustness and contribution of key technical components within \mname.

\section{Related Works}


\subsection{Clock Models}
Researchers have developed various types of clock models trained on the proxy \emph{chronological age} to estimate individuals' \emph{biological age}.
Conventionally, clinically accessible features including physiological measurements~\cite{lima2021deep}, physical functions~\cite{wagner2016biomarkers,nofuji2016associations} and laboratory biomarkers~\cite{mitnitski2015age} have been proposed to develop \emph{phenotypic clock models} over the last several decades. Machine learning models (\textit{e.g.} linear regressions, tree models) are widely used for phenotypic clocks.
In parallel, \emph{epigenetic clock models} have emerged, leveraging DNA methylation data to capture aging-related molecular changes with high sensitivity and precision. Due to the high dimensionality of methylation profiles, they typically cover hundreds of thousands of CpG sites. Early models~\cite{hannum2013genome,horvath2013dna} often rely on manually selecting a subset of age-associated CpGs through statistical criteria or biological relevance. With the advancements in deep neural networks (DNN), researchers have started to explore DNN-based epigenetic clock models~\cite{de2022pan}. 
More recently, large-scale pre-trained foundation models~\cite{de2024cpgpt,ying2024methylgpt} have demonstrated further improvements in generalizability and predictive power for epigenetic age estimation. 

\subsection{Multimodal Machine Learning Models}
Fusion~\cite{lu2023MuG} plays a central role in multimodal machine learning by determining how to effectively integrate heterogeneous data modalities. In the context of biological age estimation, combining epigenetic features (such as DNA methylation data) with phenotypic features (including clinical and demographic variables) has shown promise in capturing complementary aspects of aging~\cite{levine2018epigenetic}. Fusion strategies are typically categorized based on when the integration occurs: early fusion~\cite{lu2025early} concatenates raw features from each modality before joint modeling, enabling the learning of unified representations. Late fusion~\cite{erickson2020autogluon}, on the other hand, processes each modality independently and combines their outputs at the decision level, providing modularity and resilience to missing data. 
\section{Study Design}
\label{sec:study_design}

\subsection{Data Source and Cohort Selection}
We use data from LinkedOmics~\cite{vasaikar2018linkedomics} which provides multi-omics data and electronic health records integrating from TCGA~\cite{weinstein2013tcga} and CPTAC~\cite{ellis2013cptac}. In this study, we select 8 cancer types for internal training and validation, and 4 distinct cancer types for external evaluation, to assess the generalizability of our models across cohorts. Table~\ref{tab:patient_cohorts} summarizes the number of subjects and chronological age characteristics of each cancer cohort, where subjects with both epigenetic and phenotypic features presented are kept.
The constructed datasets are publicly available at \url{https://dx.doi.org/10.6084/m9.figshare.29151062}. We use the cancer type abbreviation following the LinkedOmics convention.

\begin{table}[htbp!]
\centering
\vspace{-0.2cm}
\caption{Characteristics of patient cohorts in our dataset.}
\label{tab:patient_cohorts}
\vspace{-0.1cm}
\begin{tabular}{lccc}
\toprule
\textbf{Cancer Type} & \textbf{Dataset} & \textbf{\# Subjects} & \textbf{Avg Age (SD)} \\
\midrule
BRCA              & Internal & 769  & 58.2 (13.1) \\
COADREAD             & Internal & 389  & 64.4 (13.0) \\
HNSC & Internal & 527  & 60.9 (11.9) \\
LUAD                 & Internal & 429  & 65.0 (10.2) \\
LUSC         & Internal & 360  & 67.6 (8.7)  \\
KIPAN                     & Internal & 655  & 60.5 (12.6) \\
STES      & Internal & 571  & 64.3 (11.2) \\
THCA      & Internal & 503  & 47.3 (15.8) \\
\midrule
BLCA        & External & 411  & 68.1 (10.6) \\
PAAD         & External & 184  & 64.8 (11.0) \\
SKCM            & External & 105  & 64.7 (13.9) \\
TGCT         & External & 134  & 32.0 (9.3) \\
\bottomrule
\end{tabular}
\vspace{-0.2cm}
\end{table}

Table~\ref{tab:feat} provides an overview of the epigenetic and phenotypic features used for biological age estimation. The epigenetic features are collected from DNA methylation (DNAm) data measured at the CpG-site level in tumor samples, using the Illumina HM450K platform. The phenotypic features include the subject’s biological sex and a set of clinical-pathological variables (e.g., radiation therapy status and pathological stage). There is a small discrepancy in the dimensionality of the DNAm data between the internal and external cohorts due to the batch effect in the original data collection process~\cite{vasaikar2018linkedomics,weinstein2013tcga}.

\begin{table}[htbp!]
\centering
\vspace{-0.2cm}
\caption{Overview of epigenetic and phenotypic features.}
\label{tab:feat}
\vspace{-0.1cm}
\begin{tabular}{c|c}
\toprule
\textbf{Feature Name} & \textbf{Dimension/Value Set}\\
\midrule
\multicolumn{2}{l}{Epigenetic Features} \\
DNAm (Internal) & 334,362\\
DNAm (External) & 334,022\\
\hline
\multicolumn{2}{l}{Phenotypic Features}\\
Biological Sex & male; female\\
Radiation Therapy & yes; no\\
Pathologic Overall Stage & is; i; ii; iii; iv\\
Pathologic T Stage & t1; t2; t3; t4\\
Pathologic N Stage & n0; n1; n2; n3\\
Pathologic M Stage & m0; m1\\
\bottomrule
\end{tabular}    
\vspace{-0.3cm}
\end{table}

\subsection{Problem Definition}
Given the constructed dataset, we mathematically define our problem and data as follows:
 \begin{definition}[Biological Age Estimation]
The model $f_\theta(\cdot)$ is expected to use $i$-th cancer patient's DNA methylation data of tumor sample at CpG-site level $\mathbf{X}_i$ and clinical-pathological features $\mathbf{C}_i$, to estimate the cancer patient's biological age $\hat{y}_i=f_\theta(\mathbf{X}_i, \mathbf{C}_i)$ where $\hat{y}_i\in\mathbb{R}_{>0}$.
\end{definition}

\begin{definition}[DNA Methylation Data]
One patient's array-based DNA methylation profile of tumor sample at CpG-site level $\mathbf{X}_i$ is represented by a dense vector $\mathbf{X}_i\in [-0.5,+0.5]^{m}$, with each element $x_j\in [-0.5, +0.5]$ corresponding to the methylation level of the $j$-th CpG site expressed as $\beta_\text{value} -0.5$. 
\end{definition}

\begin{definition}[Clinical-Pathological Features]
One patient's clinical-pathological features $\mathbf{C}_i$ are represented by a vector $\mathbf{C}_i=[c_1,c_2,\dots,c_d]$, where each element $c_j$ is a categorical variable corresponding to the $j$-th clinical or pathological characteristic. 
\end{definition}

\section{Methods}

\begin{figure*}[htbp!]
  \centering 
  \includegraphics[width=0.9\textwidth]{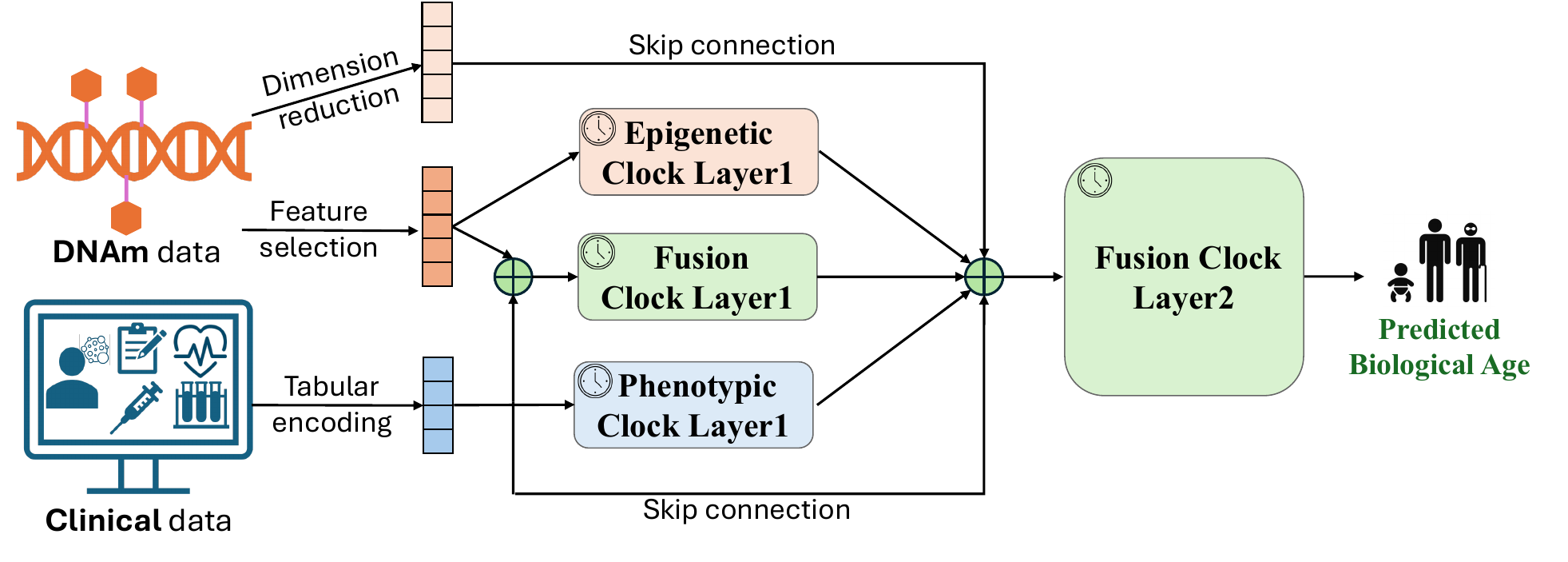}
  \vspace{-0.4cm}
  \caption{Model overview of \mname.}
  \vspace{-0.3cm}
  \label{fig:model_framework}
\end{figure*}

To fully leverage the strengths of the two data modalities for biological age estimation in cancer patients, we propose \mname (\textbf{\underline{Epi}}genetic and phenotypi\textbf{\underline{C}} based biological \textbf{\underline{Age}} prediction), a multimodal, multi-layer stacked ensemble~\cite{erickson2020autogluon,lu2025early} model. An overview of the model architecture is shown in Figure~\ref{fig:model_framework}.
In a nutshell, \mname consists of two layers of base clock models, each designed to process a distinct data modality for estimating the patient's biological age. Specifically in the first layer, one epigenetic clock model is trained on DNAm data, one phenotypic clock model is trained on clinical features, and one fusion clock model is trained on fused DNAm and clinical features. 
In the final layer, another meta-level fusion model integrates the predictions from all base models, augmented by a skip connection~\cite{he2016deep} from the raw features, to produce the final biological age estimation. This architecture effectively captures complementary signals from both omics and clinical data sources.
We provide the source code and additional technical details of \mname at \url{https://github.com/GAIN-Lab/EPICAGE}.

\subsection{Epigenetic Clock}
As Fig.~\ref{fig:model_framework} shows, the epigenetic clock resides in the first layer of \mname, which estimates the subject's age $\hat{y}^{epi}_i$ purely based on the input DNAm data $\mathbf{X}_i\in \mathbb{R}^m$. 

\subsubsection{Feature Selection}
\label{ssec:feat_select}
Following the common practice to handle the high dimensional DNAm data ($m \approx 334K$), we design a \textit{two-step feature selection module} $\Phi\mathbb{R}^m \rightarrow \mathbb{R}^k$ to select a subset of CpG indices
\begin{equation}
    \mathcal{S} \subseteq \{1, \dots, m\}, \lvert \mathcal{S} \rvert=k \ll m.
\end{equation}
The \textit{first feature selection step} is correlation filtering. For each CpG site $j=1,\dots,m$, we compute its Spearman rank correlation with the target chronological age. The top $k_{s_0}$ (e.g. we use $k_{s_0} = 2000$) with the highest absolute correlation $|\rho|$ is kept to form the initial subset:
\begin{equation}
S_0 = \text{Top-}k_{s0}\left(|\rho_j|\right)_{j=1}^{m}
\end{equation}


\noindent The \textit{second feature selection step} uses BorutaShap~\cite{ekeany2020borutashap}, a wrapper algorithm combining Boruta~\cite{kursa2010boruta} with SHAP~\cite{lundberg2017shap}, to refine the initial subset $\mathcal{S}_0$ and produce the final feature set $\mathcal{S}=BorutaSHAP(\mathcal{S}_0)$. While the first-step Spearman filtering captures CpGs with monotonic age correlations, BorutaShap further selects features that are predictive and biologically meaningful, capturing nonlinear patterns and interactions in aging-related methylation.


\subsubsection{Clock Model Implementation}
After the feature selection module $\Phi$, we obtain a lower-dimensional DNAm matrix $\mathbf{X}_{:, \mathcal{S}} = \Phi(\mathbf{X})$. We implement the clock model using a pre-trained tabular foundation model TabPFN~\cite{hollmann2025tabpfn}, which is specifically designed for small-to-medium-sized tabular data. TabPFN is well-suited for our task as it enables in-context learning by conditioning directly on the dataset without requiring costly gradient-based fine-tuning:
\begin{equation}\label{eq:epi_age}
    \hat{\mathbf{y}}^{epi}_i=f^{epi}(\mathbf{X}_{:,\mathcal{S}}^{train},\mathbf{y}^{train},\mathbf{X}_{i,\mathcal{S}}),
\end{equation}
where the TabPFN model $f^{epi}$ uses the training set $(\mathbf{X}_{:,\mathcal{S}}^{train},\mathbf{y}^{train})$ as context to predict the biological ages $\hat{\mathbf{y}}^{epi}_i$ for the $i$-th test sample $\mathbf{X}_{i,\mathcal{S}}$.

\subsection{Phenotypic Clock}
Similar to the epigenetic clock, the phenotypic clock $f_{\theta}^{phe}$ also resides in the first layer of \mname. $f_{\theta}^{phe}$ takes $i$-th subject's clinical-pathological variables $C_i$ and estimates the biological age $\hat{y}^{phe}_i$. As described in Sec.~\ref{sec:study_design}, the clinical-pathological variables are mostly categorical and boolean data. Following the common practice~\cite{erickson2020autogluon,lu2023MuG}, we apply ordinal encoding to map each categorical feature into monotonically increasing integers.
As the dimensionality $d$ for clinical variables is modest, we then directly employ another TabPFN $f^{phe}$ for the age estimation:
\begin{equation}\label{eq:pheno_age}
    \hat{\mathbf{y}}^{phe}_i=f^{phe}(\mathbf{C}^{train},\mathbf{y}^{train},\mathbf{C}_i).
\end{equation}
It is worth noting that both the epigenetic and phenotypic clocks can be implemented using alternative models such as ElasticNet~\cite{zou2005elasticnet}. IWe implement two variants of \mname (\mname-ElasticNet and -TabPFN) and evaluate them in experiments to validate our design choices.

\subsection{Fusion Clocks}
We employ two fusion clocks in layer 1 and layer 2.
\subsubsection{Layer 1 Fusion Clock}
As illustrated in Fig.~\ref{fig:model_framework}, the layer 1 fusion clock model estimates the $i$-th biological age from the concatenation $\oplus$ of the inputs of layer 1 epigenetic clock and phenotypic clock. To keep \mname a unified framework, we still implement fusion clock via TabPFN, thus
\begin{equation}\label{eq:fuse_age}
    \hat{\mathbf{y}}^{fuse}_i=f^{fuseL1}(\mathbf{X}^{train}_{:,\mathcal{S}}\oplus \mathbf{C}^{train},\mathbf{y}^{train},\mathbf{X}_{i,\mathcal{S}}\oplus\mathbf{C}_i).
\end{equation}

\subsubsection{Layer 2 Fusion Clock}
After all layer 1 clock models have been fit, \mname ensembles their estimated biological ages along with a skip connection from the raw features to produce the final estimation using a layer 2 fusion clock model. The concatenated input can be formally denoted as
\begin{equation}
\mathbf{z}_i=\hat{y}^{epi}_i\oplus\hat{y}^{phe}_i\oplus\hat{y}^{fuse}_i\oplus\mathbf{\tilde{X}}_i\oplus \mathbf{C}_i,
\end{equation}
where $\hat{y}^{epi}_i,\hat{y}^{phe}_i,\hat{y}^{fuse}_i$ are obtained by Equation~\eqref{eq:epi_age},~\eqref{eq:pheno_age},~\eqref{eq:fuse_age}, respectively. $\mathbf{\tilde{X}}_i$ denotes a dimension-reduced DNAm profile. 
We choose to use the dimension-reduced DNAm $\mathbf{\tilde{X}}_i$ instead of the feature-selected DNAm $\mathbf{X}_{i,\mathcal{S}}$ (used in Equation~\eqref{eq:epi_age}) to offer complementary views of the methylation profile. While feature selection emphasizes predictive markers, dimensionality reduction (e.g., via PCA or autoencoders) captures latent structures and global patterns in the data. This perspective is analogous to multi-head attention in Transformers, where each head attends to different aspects of the input. 
Specifically, we opt for Principal Component Analysis (PCA) for obtaining $\mathbf{\tilde{X}}_i\in \mathbb{R}^{r}, r\ll m$: $\mathbf{\tilde{X}}_i = \mathbf{X}_i \mathbf{W}_{\text{PCA}}$.
Similar to previous clock models, we implement the layer 2 fusion clock using TabPFN, thus
$\hat{\mathbf{y}}_i=f^{fuseL2}(\mathbf{z}^{train},\mathbf{y}^{train},\mathbf{z}_i)$.


\section{Experiments}
\label{sec:exp}
In this section, we evaluate our proposed \mname, focusing on the following four research questions:
\textbf{RQ1}: How accurate is the proposed model compared to baseline methods?
\textbf{RQ2}: Can the model effectively identify aging-related CpG sites?
\textbf{RQ3}: What are the clinical implications of the \mname-derived biological age? 
\textbf{RQ4}: What insights can be gained through error analysis and ablation studies?

\subsection{Compared Baseline Models}
We compare \mname with epigenetic and phenotypic clock models. \textbf{Epigenetic clock} include: 
(a.1) \textbf{\textit{Horvath}}~\cite{horvath2013dna}: A multi-tissue elastic net clock using 353 CpGs from 7,844 samples across 51 tissues to estimate biological age.
(a.2) \textbf{\textit{Hannum}}~\cite{hannum2013genome}: A blood-based elastic net clock trained on 656 samples selecting 71 CpGs as age predictors.
(a.3) \textbf{\textit{PhenoAge}}~\cite{levine2018epigenetic}: An elastic net clock trained to predict phenotypic age from 20,169 blood samples using 513 CpGs.
(a.4) \textbf{\textit{YingCausAge}}~\cite{ying2024causeage}: A causality-aware elastic net clock leveraging CpG-level causality scores from 2,664 blood methylation samples.
(a.5) \textbf{\textit{AltumAge}}~\cite{de2022pan}: A deep neural network-based pan-tissue clock using 20,318 CpGs from 8,050 samples across multiple platforms.
We follow common practice by using publicly released model weights for most epigenetic clocks, except for AltumAge, which is fine-tuned on our internal dataset.
For \textbf{phenotypic clocks}, we benchmark the following popular machine learning models:
(b.1) \textbf{\textit{Linear Regression}}: Models age as a weighted linear combination of features.
(b.2) \textbf{\textit{Random Forest}}~\cite{breiman2001random}: A bagging-based ensemble of decision trees that averages predictions for robustness.
(b.3) \textbf{\textit{XGBoost}}~\cite{chen2016xgboost}: A gradient boosting algorithm that sequentially builds trees to minimize residual errors.
(b.4) \textbf{\textit{LightGBM}}~\cite{ke2017lightgbm}: A computationally efficient, leaf-wise gradient boosting framework using histogram-based splits.
(b.5) \textbf{\textit{tabNN}}~\cite{erickson2020autogluon}: A neural network optimized for tabular data with automatic feature scaling and regularization.

\subsection{Evaluation of Age Estimation Performance (RQ1)}
To address RQ1, we conduct two sets of evaluations to assess our model's performance in estimating chronological age. Since chronological age serves as a widely accepted proxy for biological age, and our models are trained on that, evaluating prediction errors provides a reliable measure of the accuracy of our proposed biological clock model.
For the \textit{internal evaluation}, we perform five-fold cross-validation across eight internal cancer cohorts (see Table~\ref{tab:patient_cohorts} for cohort details).
For the \textit{external evaluation}, we re-train the models on all internal cohort patients and assess their generalizability on four additional cancer cohorts that were not included in the training process.

\begin{table*}[htbp!]
\centering
\caption{Chronological age estimation results.}
\vspace{-0.2cm}
\begin{subtable}[t]{0.58\textwidth}
\caption{Internal eight cancer cohorts of 4,203 patients (mean ± SD).}
\label{tab:internal_eval}
\vspace{-0.1cm}
\begin{tabular}{lccc}
\toprule
\textbf{Model} & \textbf{RMSE} & \textbf{MAE} & \textbf{R\textsuperscript{2}} \\
\midrule
\multicolumn{4}{l}{\textit{Epigenetic Clock Models}}\\
\quad Horvath & 22.15 ± 0.80 & 17.19 ± 0.48 & -1.67 ± 0.28 \\
\quad Hannum & 19.55 ± 0.40 & 15.39 ± 0.31 & -1.07 ± 0.16 \\
\quad PhenoAge & 37.04 ± 1.74 & 27.96 ± 1.26 & -6.47 ± 0.94 \\
\quad YingCausAge & 23.22 ± 0.86 & 17.87 ± 0.79 & -1.93 ± 0.25 \\
\quad AltumAge& 10.49 ± 0.69 & 8.23 ± 0.58 & 0.40 ± 0.08\\
\midrule
\multicolumn{4}{l}{\textit{Phenotypic Clock Models}} \\
\quad LinearRegression & 12.87 ± 0.29 & 10.40 ± 0.24 & 0.10 ± 0.02 \\
\quad RandomForest & 13.09 ± 0.31 & 10.41 ± 0.19 & 0.07 ± 0.04 \\
\quad XGBoost & 12.50 ± 0.34 & 10.01 ± 0.23 & 0.15 ± 0.03 \\
\quad LightGBM & 12.49 ± 0.32 & 10.03 ± 0.22 & 0.16 ± 0.02 \\
\quad tabNN & 12.41 ± 0.33 &  9.91 ± 0.23 & 0.17 ± 0.03 \\
\midrule
\multicolumn{4}{l}{\textit{Fusion Clock Models}}\\
\quad \textbf{\mname}-ElasticNet & \underline{9.99} ± 0.30 & \underline{8.01} ± 0.18 & \underline{0.46} ± 0.02 \\
\quad \textbf{\mname}-TabPFN & \textbf{8.16} ± 0.18 & \textbf{6.46} ± 0.13 & \textbf{0.64} ± 0.02 \\
\bottomrule
\end{tabular}
\end{subtable}
\hfill
\begin{subtable}[t]{0.4\textwidth}
\centering
\caption{External four cancer cohorts of 834 patients.}
\label{tab:external_eval}
\vspace{-0.1cm}
\begin{tabular}{lccc}
\toprule
\textbf{Model} & \textbf{RMSE} & \textbf{MAE} & \textbf{R\textsuperscript{2}} \\
\midrule
\multicolumn{4}{l}{\textit{Epigenetic Clock Models}}\\
\quad Horvath & 23.58 & 18.46 & -0.96 \\
\quad Hannum & 20.81 & 16.99 & -0.52 \\
\quad PhenoAge & 36.08 & 27.62 & -3.58 \\
\quad YingCausAge & 23.06 & 17.81 & -0.87 \\
\quad AltumAge & 15.34 & 11.61 & 0.17 \\
\midrule
\multicolumn{4}{l}{\textit{Phenotypic Clock Models}}\\
\quad LinearRegression & 17.17 & 13.36 & -0.04 \\
\quad RandomForest     & 17.44 & 13.67 & -0.07 \\
\quad XGBoost          & 16.57 & 13.08 & 0.03 \\
\quad LightGBM         & 16.62 & 13.14 & 0.03 \\
\quad tabNN            & 17.18 & 13.33 & -0.04 \\
\midrule
\multicolumn{4}{l}{\textit{Fusion Clock Models}}\\
\quad \textbf{\mname}-ElasticNet  & \underline{12.62} & \underline{10.28} & \underline{0.44} \\
\quad \textbf{\mname}-TabPFN  & \textbf{9.11} & \textbf{7.18} & \textbf{0.71} \\
\bottomrule
\end{tabular}
\end{subtable}
\vspace{-0.3cm}
\end{table*}

On the internal test set (Table~\ref{tab:internal_eval}), EPICAGE-TabPFN achieves substantial gains over existing clocks, with a 63.72\% reduction in RMSE and a 62.72\% reduction in MAE compared to the average of epigenetic clocks, along with an absolute increase of 2.79 in $R^2$. Relative to phenotypic clocks, RMSE and MAE are reduced by 35.61\% and 36.37\%, respectively, with an $R^2$ improvement of 0.51.
EPICAGE-ElasticNet also demonstrates strong performance, with a 55.58\% reduction in RMSE and a 53.77\% reduction in MAE over epigenetic clocks, and corresponding decreases of 21.16\% and 21.10\% over phenotypic clocks. The absolute gains in $R^2$ are 2.61 and 0.33, respectively.

On the external test set (Table~\ref{tab:external_eval}), these trends persist. EPICAGE-TabPFN achieves a 61.68\% reduction in RMSE and a 61.18\% reduction in MAE compared to epigenetic clocks, with an $R^2$ increase of 1.86. When compared to phenotypic clocks, the model yields 46.40\% and 46.08\% reductions in RMSE and MAE, respectively, and a 0.73 improvement in $R^2$. EPICAGE-ElasticNet delivers consistent improvements as well, with RMSE and MAE reduced by 46.92\% and 44.43\% over epigenetic clocks, and by 25.75\% and 22.80\% over phenotypic clocks, accompanied by $R^2$ increases of 1.59 and 0.46, respectively.
Among all baseline models, AltumAge demonstrates relatively strong performance due to its fine-tuning on cancer-specific data. Nonetheless, it still underperforms our models. These results underscore the benefit of multimodal feature integration and ensemble modeling, even when compared against specialized deep learning clocks trained on similar cancer cohorts.

\subsection{Analysis of Aging-Related CpG Sites (RQ2)}

To address RQ2, we conduct four analyses to examine whether our model captures biologically meaningful and potentially cancer-specific aging signals. 
\mname uses a two-stage feature selection pipeline for CpG sites on each training fold. Each fold yields approximately 290 CpG sites, with 57 CpGs consistently selected in all five folds.
All analyses are conducted on the 57 CpGs.

\subsubsection{Validation through Overlap with Established Epigenetic Clocks}  
Among the CpG sites selected by our \mname, 8 overlap with those included in the widely used Horvath, Hannum, PhenoAge, and YingCausAge epigenetic clocks, suggesting that our approach captures established, biologically relevant methylation signals associated with aging. 
Notably, \textit{cg22736354} appears in all nine tissue-specific clocks reported by \cite{choi2019development}, while \textit{cg23606718}, annotated to the \textit{ELOVL2} gene, is a robust and widely validated marker of aging across tissues and populations~\cite{gopalan2019dna}.

\subsubsection{Comparative Evaluation of \mname-Selected CpGs Against Existing Epigenetic Clocks}
To assess whether the CpG sites identified by our model capture cancer-relevant aging signals recognized by existing clocks, we conduct a comparative analysis of age prediction performance on both internal and external cancer datasets.
\begin{itemize}
\item CpGs from existing epigenetic clocks;
\item The full set of 57 CpGs selected by our \mname;
\item A subset of 49 CpGs from our model after excluding the 8 sites overlapping with existing clocks (Horvath, Hannum, PhenoAge, YingCausAge).
\end{itemize}
As shown in Table~\ref{stab:cpg_effect_internal}, the model trained on our 57 CpGs achieves comparable performance to models trained on existing clock CpGs. Notably, even after removing the 8 overlapping sites, the remaining 49 CpGs maintains similar predictive accuracy. In Table~\ref{stab:cpg_effect_external}, which reports results on external cancer cohorts, our model achieves the second-best performance overall.
These findings suggest that our \mname model identifies additional aging-associated CpG sites that may capture cancer-specific aging signals not included in existing epigenetic clocks.

\begin{table*}[htbp!]
\centering
\caption{Ablation study on selected CpG sites for chronological age estimation. All variants are trained with ElasticNet.}
\label{tab:cpg_effect}
\vspace{-0.2cm}
\begin{subtable}[t]{0.58\textwidth}
\centering
\caption{Internal eight cancer cohorts (mean ± SD).}
\label{stab:cpg_effect_internal}
\vspace{-0.1cm}
\begin{tabular}{llccc}
\toprule
\textbf{CpG Source} & \textbf{\#CpGs}& \textbf{RMSE} & \textbf{MAE} & \textbf{R\textsuperscript{2}} \\
\midrule
Horvath & 353 & 11.10 ± 0.35 & 8.93 ± 0.27 & 0.33 ± 0.02 \\
Hannum & 71 & \underline{10.95} ± 0.24 & \underline{8.84} ± 0.22 & \underline{0.35} ± 0.03 \\
PhenoAge & 514 & 10.98 ± 0.39 & \underline{8.81} ± 0.29 & \underline{0.35} ± 0.03 \\
YingCausAge & 586 & 11.25 ± 0.37 & 9.06 ± 0.24 & 0.31 ± 0.03 \\
\midrule
\mname & 57 & \textbf{10.69} ± 0.30 & \textbf{8.59} ± 0.20 & \textbf{0.38} ± 0.03 \\
\quad $-$ Overlap & 49 & 11.05 ± 0.33 & 8.87 ± 0.22 & 0.34 ± 0.03 \\
\bottomrule
\end{tabular}
\end{subtable}
\hfill
\begin{subtable}[t]{0.4\textwidth}
\centering
\caption{External four cancer cohorts.}
\label{stab:cpg_effect_external}
\vspace{-0.1cm}
\begin{tabular}{llccc}
\toprule
\textbf{CpG Source} & \textbf{\#CpGs}& \textbf{RMSE} & \textbf{MAE} & \textbf{R\textsuperscript{2}} \\
\midrule
Horvath &  353 & 13.27 & \underline{10.66} & 0.38 \\
Hannum & 71 & \textbf{12.29} & \textbf{10.09} & \textbf{0.47} \\
PhenoAge & 514 & 13.48 & 10.82 & 0.36 \\
YingCausAge & 586 & 14.53 & 11.95 & 0.26 \\
\midrule
\mname & 57 & \underline{13.19} & 10.71 & \underline{0.39} \\
\quad $-$ Overlap & 49 & 14.38 & 11.47 & 0.27 \\
\bottomrule
\end{tabular}    
\end{subtable}
\vspace{-0.2cm}
\end{table*}

\subsubsection{Supporting Evidence from Published CpG-Trait Associations}
To further assess the biological relevance of the CpG sites selected by our model, we query all 57 CpGs against the EWAS Atlas, a curated knowledge base of epigenome-wide association studies~\cite{li2019ewas}. This resource catalogs published associations between CpG methylation and diverse traits or diseases. Our query reveals that 50 out of the 57 CpG sites have been reported to be associated with cancer or aging. These results provide independent evidence that our selected CpGs are robustly linked to aging and cancer processes as documented in the literature.

\subsubsection{Functional Enrichment of Genes Near Selected CpGs} 
To explore the potential biological pathways represented by the CpG sites identified by our model, we conduct a KEGG pathway enrichment~\cite{garcia2022functional} analysis on genes proximal to the 57 CpG sites. The analysis reveals significant enrichment in pathways relevant to aging and cancer. The ``\textit{Longevity Regulating Pathway}'' (genes: \textit{SOD2}, \textit{IRS2}; $p=0.002$) and ``\textit{MicroRNAs in Cancer}'' (genes: \textit{TP63}, \textit{IRS2}; $p=0.016$) pathways were among the most enriched. These findings suggest that the methylation changes captured by our model may influence gene networks and regulatory mechanisms that are critically involved in both aging and oncogenesis.

\subsection{Assessing Clinical Utility of \mname-Derived Biological Age (RQ3)}

\begin{table}[htbp!]
\centering
\caption{Cox survival analysis results of all patients.}
\label{tab:cox_overall}
\vspace{-0.1cm}
\begin{tabular}{lccc}
\toprule
\textbf{Model} & \textbf{HR (per 5y)} & \textbf{95\% CI} & \textbf{p-value} \\
\midrule
\multicolumn{4}{c}{\textit{Internal Eight Cancer Cohorts}} \\
Horvath            & \cellcolor{blue!20} 0.953 & (0.938, 0.968) & $\mathbf{<0.01}$ \\
Hannum             & \cellcolor{blue!10} 0.988 & (0.970, 1.006) & 0.177 \\
\mname     & \cellcolor{red!15} 1.067 & (1.018, 1.119) & $\mathbf{<0.01}$ \\
\hline
\multicolumn{4}{c}{\textit{External Four Cancer Cohorts}} \\
Horvath            & \cellcolor{red!10} 1.042 & (1.018, 1.067) & $\mathbf{<0.01}$ \\
Hannum             & \cellcolor{red!15} 1.067 & (1.031, 1.104) & $\mathbf{<0.01}$ \\
\mname     & \cellcolor{red!30} 1.113 & (1.027, 1.207) & $\mathbf{<0.01}$ \\
\bottomrule
\end{tabular}
\vspace{-0.3cm}
\end{table}

To investigate the clinical utilities (RQ3) of our model-derived biological aging, we conduct survival analysis to examine the association between age acceleration and patient outcomes~\cite{levine2018epigenetic}. Survival analysis is particularly relevant in the cancer population, where prognosis and mortality risk are key concerns~\cite{yu2020epigenetic}. Specifically, we employ the Cox proportional hazards model~\cite{cox1972regression} to evaluate whether deviations in biological age, as captured by age acceleration, are predictive of overall survival independent of chronological age.
For this analysis, we assess the relationship between \textbf{age acceleration} ($\text{AA}$) and cancer survival, while adjusting for chronological age and biological sex.  Following conventions in survival analysis, we divide ($\text{AA}$) by 5 to express the hazard ratio (HR) per 5-year increase in age acceleration.
Table~\ref{tab:cox_overall} presents the Cox survival analysis~\cite{cox1972regression} results for both internal and external cancer cohorts. As shown, our \mname model reveals a significant positive association between age acceleration and mortality risk ($p < 0.01$). Specifically, each 5-year increase in age acceleration is associated with a 6.7\% and 11.3\% \emph{increase} in the hazard of mortality in the internal and external cohorts, respectively.
In comparison, Horvath’s age acceleration exhibits a significant \emph{inverse} association with mortality in the internal cohorts, where each 5-year increase in age acceleration is associated with a 4.7\% \emph{reduction} in the hazard of mortality. In contrast, Hannum’s age acceleration shows a \emph{non-significant} inverse association with mortality in the internal cohorts. For the external cohorts, both Horvath’s and Hannum’s models present a similar pattern to our \mname model.


\subsection{Error Analysis and Ablation Study (RQ4)}


\begin{figure}[H]
  \centering
  \vspace{-0.2cm}
  \includegraphics[width=0.4\textwidth]{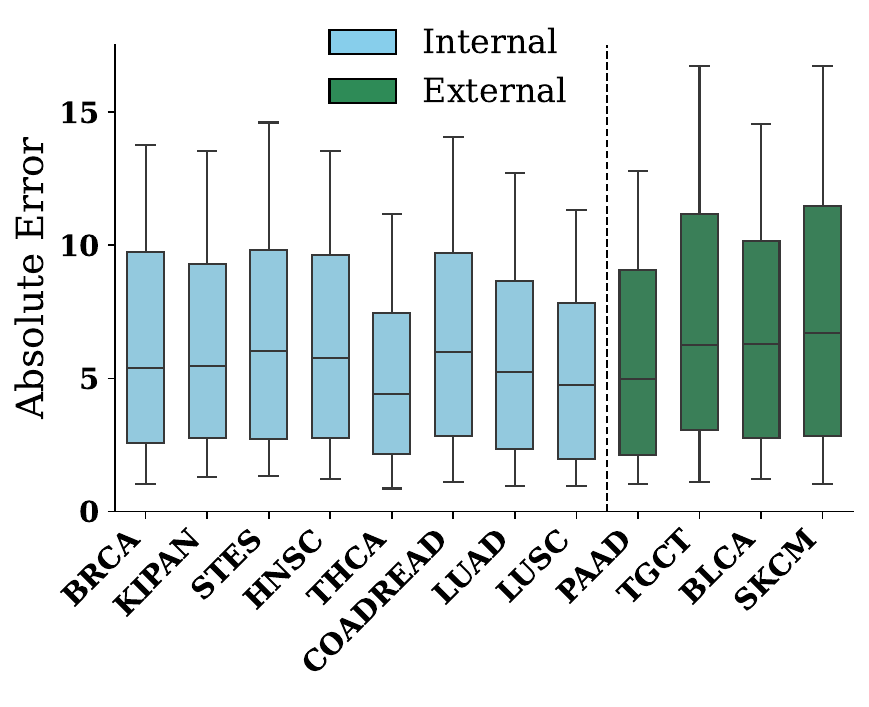}
  \vspace{-0.2cm}
  \caption{Absolute Error of Age Prediction by Cancer Type.}
\label{fig:absoluteError_cancer}
\vspace{-0.2cm}
\end{figure}

\subsubsection{Error Analysis}
To further understand the performance of our age prediction model across different clinical contexts and cancer cohorts, we conducted an error analysis. 
As shown in Figure~\ref{fig:absoluteError_cancer}, the absolute prediction error across cancer types in the internal dataset remains relatively low and stable, with median errors typically around 5–6 years. In contrast, the external validation dataset shows higher and more variable errors, with BLCA, TGCT, and SKCM exhibiting median errors exceeding 6 years and broader error distributions. This pattern indicates that while the model achieves consistent performance on internal data, its generalization to external cohorts is more limited. 




\begin{table}[H]
\centering
\caption{Ablation study results.}
\label{tab:ablation}
\vspace{-0.1cm}
\begin{tabular}{lccc}
\toprule
\textbf{Method} & \textbf{RMSE}  & \textbf{MAE} & \textbf{R$^2$}  \\
\midrule
\multicolumn{4}{c}{\textit{Internal Dataset}}\\
\mname-TabPFN      & \textbf{8.16 $\pm$ 0.18} & \textbf{6.46 $\pm$ 0.13} & \textbf{0.64 $\pm$ 0.02} \\
w/o Ensemble  & 8.50 $\pm$ 0.15 & 6.71 $\pm$ 0.14 & 0.61 $\pm$ 0.02 \\
w/o Skip connection  & 8.40 $\pm$ 0.20 & 6.64 $\pm$ 0.18 & 0.62 $\pm$ 0.02 \\
\hline
\multicolumn{4}{c}{\textit{External Dataset}}\\
\mname-TabPFN       & \textbf{9.11} & \textbf{7.18} & \textbf{0.71} \\
w/o Ensemble  & 9.50  & 7.43  & 0.68 \\
w/o Skip connection    & 16.85 & 13.52 & 0.00 \\
\bottomrule
\end{tabular}
\vspace{-0.3cm}
\end{table}


\subsubsection{Ablation Study}
Our \mname framework incorporates two key technical innovations: (1) a multi-layer stacked multimodal ensemble, and (2) skip connections from raw features. To validate the necessity of each technical component, we conduct ablation studies by selectively removing these parts. Table~\ref{tab:ablation} presents results from internal five-fold cross-validation and external validation. The first row in each dataset, labeled \emph{\mname-TabPFN}, represents the full model with all components enabled. \emph{w/o Ensemble} denotes the version using only the first-layer fusion clock (see Fig.~\ref{fig:model_framework}), while \emph{w/o Skip Connection} uses the second-layer fusion clock without incorporating raw feature inputs—relying solely on the predictions from first-layer clocks.
We observe consistent performance degradation on both internal and external datasets when either component is removed, highlighting their importance. Notably, removing the skip connection leads to a substantial performance drop in external cohorts.

\section{Conclusion}
Epigenetic and phenotypic data provide complementary information for accurately estimating the biological age of cancer patients. In this paper, we propose \mname, a novel multi-layer stacked multimodal framework for effectively integrating the two distinct modalities. Extensive experiments on eight internal cancer cohorts and four external cancer cohorts demonstrate the effectiveness and robustness of \mname in biological age estimation. We further validate its biological and clinical utility through downstream analyses.
There are possible limitations of \mname. First, \mname is trained using chronological age as the surrogate target for biological age, which is a common practice in many epigenetic clock models. However, chronological age may not fully capture inter-individual variability in biological aging. Recent studies have explored alternative targets such as clinical biomarkers derived phenotypic age~\cite{levine2018epigenetic}, and frailty indices~\cite{mak2024temporal}, although these approaches require longitudinal data that may not always be available.
Second, While DNAm data are powerful for aging estimation, such high-throughput data are still not routinely collected in most clinical settings, which may limit immediate clinical translation.

In the future, we plan to extend our methodology in several directions to address the current limitations. First, we will incorporate high-resolution single-cell and multi-omics data (e.g., transcriptomics, proteomics) to capture cellular heterogeneity and identify more nuanced, cell-type-specific aging signatures in cancer. Second, we aim to explore alternative surrogate targets for model training to better reflect individual variation in biological aging, particularly in cancer contexts. Third, to enhance translational utility, we will investigate the feasibility of using more accessible biomarkers or reduced panels of DNAm sites, with the goal of developing lightweight yet accurate models suitable for real-world clinical settings.

\bibliographystyle{IEEEtran}
\bibliography{myref}

\clearpage
\appendix
\subsection{Abbreviation List of Cancer Types}\label{apssec:abbreviation_cancer}
We use the following abbreviations for the selected cancer types in this paper: 
\begin{enumerate}
    \item BRCA: Breast invasive carcinoma;
    \item COADREAD: Colorectal adenocarcinoma;
    \item HNSC: Head and neck squamous cell carcinoma;
    \item LUAD: Lung adenocarcinoma;
    \item LUSC: Lung squamous cell carcinoma;
    \item KIPAN: Pan-kidney cancer;
    \item STES: Stomach and esophageal carcinoma;
    \item THCA: Thyroid carcinoma;
    \item BLCA: Bladder urothelial carcinoma;
    \item PAAD: Pancreatic adenocarcinoma;
    \item SKCM: Skin cutaneous melanoma;
    \item TGCT: Testicular germ cell tumor.
\end{enumerate}

\subsection{Phenotypic Features Definitions}
\label{apsec:phenotypic_feats_def}

We extract the following variables from LinkedOmics~\cite{vasaikar2018linkedomics} as phenotypic features for our constructed dataset (original column names from the raw CSV file are shown in parentheses):
\begin{enumerate}
\item Biological Sex (\texttt{sex}): The biological sex of the patient.
\item Radiation Therapy (\texttt{radiation\_therapy}): A binary indicator denoting whether the patient received radiation therapy during the recorded treatment period.
\item Pathologic Overall Stage (\texttt{pathologic\_stage}): The overall pathological stage based on the TNM system, which summarizes tumor size, lymph node involvement, and metastasis status. A value of ``\emph{is}'' corresponds to Pathologic Stage 0 (carcinoma in situ), indicating that abnormal cells are present but have not yet invaded surrounding tissues.
\item Pathologic T Stage (\texttt{pathology\_T\_stage}): The "T" component of TNM staging, describing the size and local invasion of the primary tumor.
\item Pathologic N Stage (\texttt{pathology\_N\_stage}): The "N" component of TNM staging, indicating the extent of regional lymph node involvement.
\item Pathologic M Stage (\texttt{pathology\_M\_stage}): The "M" component of TNM staging, representing the presence or absence of distant metastases.
\end{enumerate}
We also incorporate several additional variables from the raw phenotypic data in our analysis:
\begin{enumerate}
\item \texttt{years\_to\_birth}: Represents the chronological age of the patient at the time of the clinical visit, used as a proxy for biological age.
\item \texttt{cancer\_type}: Identifies the specific cancer cohort to which the patient belongs.
\item \texttt{overall\_survival}: Denotes the total survival time (in days) from diagnosis or clinical visit to either death or last follow-up.
\item \texttt{status}: Indicates the patient’s vital status at the last follow-up, with 1 representing deceased and 0 representing alive.
\end{enumerate}

\subsection{Details about Feature Pre-processing}
\subsubsection{DNA Methylation Feature Pre-processing}\label{ssec:dnam_pre}
All models in this study are trained on DNA methylation data obtained from the Illumina HM450K array. The original feature identifiers include both gene annotations and CpG site IDs (e.g., \texttt{GENE\_cg12345678}). To standardize the input features, we remove the gene annotations and retain only the CpG site IDs. We also perform deduplication to eliminate redundant features. These preprocessing steps are applied consistently across all pipelines to ensure fair comparison.

\subsubsection{Model Specific Pre-processing}
For the baseline \emph{epigenetic clock models} (\textit{e.g.}, Horvath, Hannum, PhenoAge, and YingCausAge), we adopt their implementations from the open-source package \texttt{biolearn}~\cite{ying2024biolearn}. Our methylation dataset, provided by LinkedOmics, is reported in a centered format (Beta value $-$ 0.5) according to the platform; therefore, we uniformly shift the DNAm values by $+0.5$ to meet the input requirement of \texttt{biolearn}, which expects standard Beta values in the range [0,1].

For the baseline \emph{phenotypic clock models} (\textit{e.g.}, Linear Regression, Random Forest, XGBoost, LightGBM, and tabNN), we adopt their implementations from the open-source package \texttt{AutoGluon}~\cite{erickson2020autogluon}. Similar to Equation~\eqref{eq:clinical_feat} in \mname, each categorical feature is mapped into monotonically increasing integers via ordinal encoding.


\subsection{Implementation Details about Our \mname}
\subsubsection{DNAm Feature Pre-processing}
Beyond the preprocessing steps detailed in Appendix~\ref{ssec:dnam_pre}, we impute missing methylation values using the column-wise mean of the training data. Additionally, we apply Z-score normalization using the mean and standard deviation computed exclusively from the training set.

\subsubsection{Second Feature Selection Step for Epigenetic Clock: BorutaSHAP}\label{APssec:BorutaSHAP}
All steps of feature selection are conducted within the training set to avoid information leakage. 
We employ BorutaSHAP on $\mathcal{S}_0$ to obtain the final subset $\mathcal{S}$ with the following hyperparameters.
\begin{center}
\begin{tabular}{ll}
\toprule
\textbf{Hyperparameter} & \textbf{Value} \\
\midrule
model type & LightGBM Regressor \\
n estimators & 100 \\
max depth & 7 \\
learning rate & 0.05 \\
min gain to split & 1e-4 \\
min data in leaf & 5 \\
subsample & 0.8 \\
colsample bytree & 0.8 \\
importance measurement & SHAP\\
SHAP arguments & pvalue=0.05 \\
max feature count & $\leq 492$ \\
\bottomrule
\end{tabular}
\end{center}

\subsubsection{Dimension Reduction of DNAm Data for Skip-connection}
Principal Component Analysis (PCA) is performed on the training set $\mathbf{X}^{\text{train}}$, and the top $r = 400$ components are retained. These same $r = 400$ components are applied consistently during inference.

\subsubsection{Handling Extra Categorical Variable in External Dataset}

The external clinical dataset contains one additional categorical variable compared to the internal dataset. To ensure compatibility, all categorical variables were converted to the \texttt{category} type and processed using TabPFN’s internal one-hot encoder with \texttt{handle\_unknown="ignore"}~\cite{hollmann2025tabpfn}. Under this scheme, any unseen categories present in the external dataset are mapped to an all-zero vector in the one-hot representation, effectively treating them as ``unknown.'' As a result, predictions for these cases rely solely on the remaining known features, allowing models trained on the internal dataset to perform inference on the external dataset without additional modification.

\subsection{Biological Validation of Selected CpG Sites}


\subsubsection{Supporting Evidence from Published CpG-Trait Associations}\label{apssec:EWAS}
In EWAS Atlas based analysis, we find that 50 out of the 57 CpG sites have been previously reported to be associated with cancer or aging. Table~\ref{tab:CpG-summary} summarizes these CpG sites grouped by their reported trait. 
For presentation purposes, we use the broad ``TraitType'' column from EWAS analysis results (\textit{i.e.}, ``TraitType'' value equal to ``cancer'') to denote CpG sites that are statistically associated with \textit{cancer}. And we use the fine-grained ``Trait'' column (\textit{i.e.}, ``Trait'' value contains ``aging'' or ``chronological age'') from EWAS results to denote CpG sites that are associated with \textit{aging}.

\begin{table}[h!]
\centering
\caption{Summary of CpG sites by trait. }
\label{tab:CpG-summary}
\begin{tabular}{p{2cm}p{5cm}}
\toprule
Trait & CpG site \\
\midrule
Cancer & cg01341751, cg01586506, cg03181248, cg05129081, cg05454501, cg07575466, cg11418477, cg14965220, cg22153181, cg26885220\\
\hline
Aging & cg04836038, cg05289022, cg05304393, cg05404236, cg07553761, cg08928145, cg10687131, cg11705975, cg12451153, cg12934382, cg13221458, cg16832267, cg21159778, cg22736354, cg24922090, cg26792755\\
\hline
Cancer \& Aging & cg00292135, cg00590036, cg00884093, cg04875128, cg04940570, cg05207048, cg06268694, cg06458239, cg06784991, cg06933824, cg07755735, cg12920180, cg13790603, cg14780466, cg15618978, cg16015712, cg16295725, cg18795809, cg19078576, cg20809087, cg23091758, cg23606718, cg24466241, cg25352836
\\
\bottomrule
\end{tabular}
\end{table}

\subsubsection{Functional Enrichment of Genes Near Selected CpGs}\label{apssec:KEGG}

From our KEGG pathway enrichment analysis, we identify significant enrichment in pathways related to aging and cancer, including the Longevity Regulating Pathway and MicroRNAs in Cancer. Table~\ref{tab:KEGG} provides details of these enriched pathways, including the genes located near the selected CpG sites and the corresponding $p$-values.

\begin{table}[h!]
\centering
\caption{KEGG pathway analysis results.}
\label{tab:KEGG}
\begin{tabular}{ccc}
\toprule
KEGG Pathway & Genes & $p$-val \\
\midrule
Longevity regulating pathway    &  SOD2, IRS2 & 0.002\\
MicroRNAs in cancer & TP63, IRS2 & 0.016 \\
\bottomrule
\end{tabular}
\end{table}

\subsection{Age Acceleration Distribution Among Cancer Types}

\begin{figure}[H]
  \centering
  \includegraphics[width=0.4\textwidth]{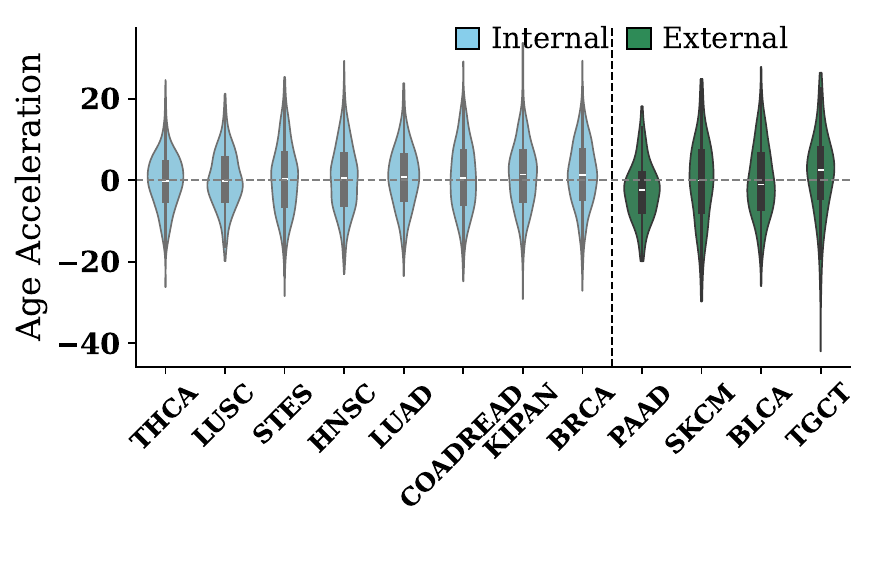}
  \vspace{-0.3cm}
  \caption{Age acceleration distribution among cancer types.}
  \label{fig:age_acceleration}
\end{figure}

Figure~\ref{fig:age_acceleration} illustrates age acceleration distributions. For the internal dataset, medians across most cancer types were close to zero, suggesting minimal bias. However, external data exhibit heterogeneous biases: TGCT showed positive median age acceleration (predicted ages higher than true), whereas PAAD and BLCA showed negative medians (predicted ages lower than true).

\subsection{More Error Analysis}\label{apssec:error_ana}
\begin{figure}[H]
  \centering
  \includegraphics[width=0.4\textwidth]{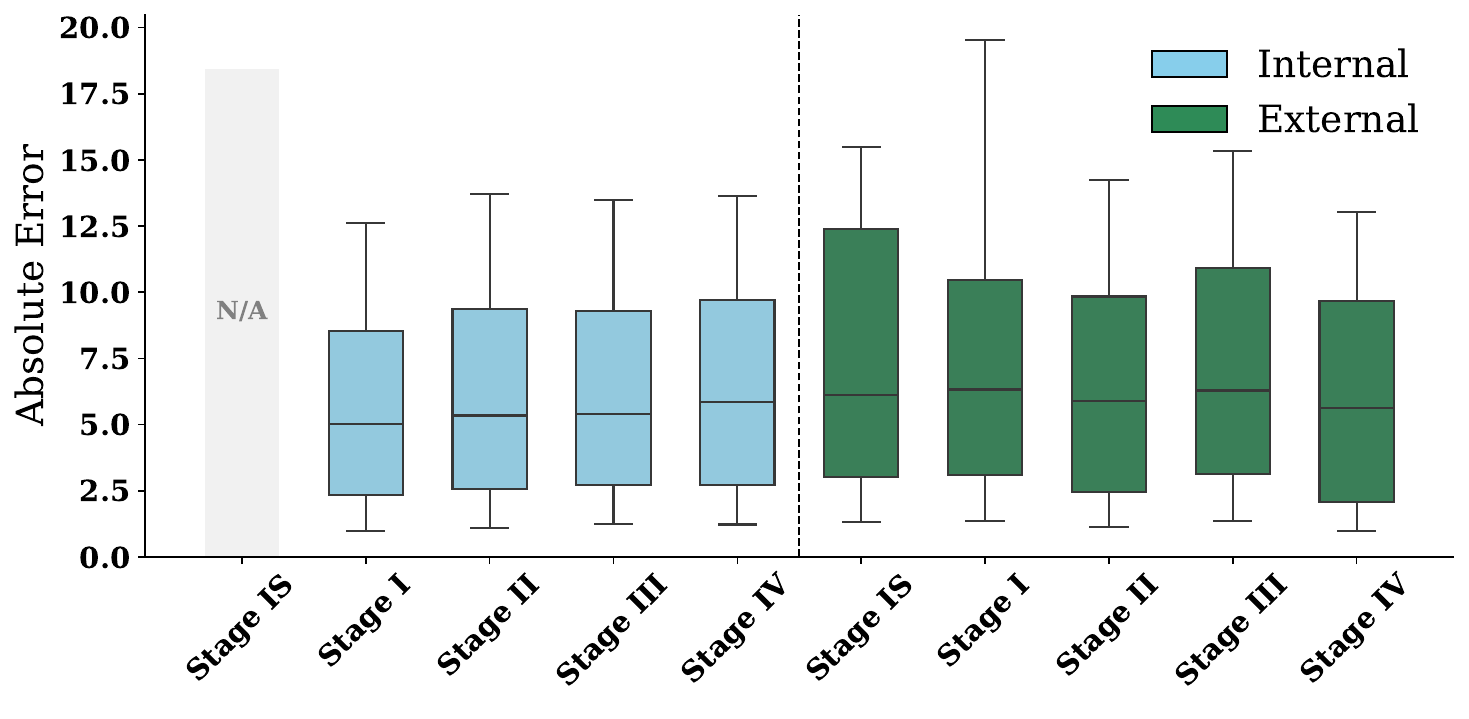}
  \caption{Absolute error of biological age prediction by stage.}
  \label{fig:absoluteError_stage}
\end{figure}


We further stratify errors by pathological stage (Figure~\ref{fig:absoluteError_stage}). For the internal dataset, median absolute errors are relatively stable across Stage~I to Stage~IV (approximately 5--6 years), showing minimal variation between stages. In contrast, the external dataset exhibits higher and more variable errors, with Stage~I showing the largest median error (about 6.3 years) and Stage~III also displaying elevated error. Rare ambiguous stages (e.g., ``I/IINOS'') are excluded due to unclear definition and only one available sample.

\end{document}